# Gas permeation through graphdiyne-based nanoporous membranes

Zhihua Zhou[1], Yongtao Tan[2,3], Qian Yang[2,3], Achintya Bera[2,3], Zecheng Xiong[4,5], Mehmet Yagmurcukardes[6], Minsoo Kim[2,3], Yichao Zou[7], Guanghua Wang[1], Artem Mishchenko[2,3], Ivan Timokhin[2,3], Canbin Wang[1], Hao Wang[1], Chongyang Yang[1], Yizhen Lu[1], Radha Boya[2,3], Honggang Liao[1], Sarah Haigh[7], Huibiao Liu[4,5], Francois M. Peeters[8], Yuliang Li[4,5], Andre K. Geim[2,3], Sheng Hu[1]

[1] State Key Laboratory of Physical Chemistry of Solid Surfaces, Collaborative Innovation Center of Chemistry for Energy Materials (iChEM), College of Chemistry and Chemical Engineering, Xiamen University, Xiamen 361005, P. R. China.
[2] Department of Physics and Astronomy, University of Manchester, Manchester M13 9PL, UK.
[3] National Graphene Institute, University of Manchester, Manchester M13 9PL, UK.
[4] Beijing National Laboratory for Molecular Sciences (BNLMS), CAS Research/Education Center for Excellence in Molecular Sciences, Institute of Chemistry, Chinese Academy of Sciences, Beijing 100190, P. R. China.
[5] University of Chinese Academy of Sciences, Beijing 100049, P. R. China.
[6] Department of Photonics, Izmir Institute of Technology, 35430 Izmir, Turkey.
[7] Department of Materials, University of Manchester, Manchester M13 9PL, UK.
[8] Department of Physics, University of Antwerp, Groenenborgerlaan 171, B-2020 Antwerp, Belgium.

**Nanoporous membranes based on two dimensional materials are predicted to provide highly selective gas transport in combination with extreme permeability. Here we investigate membranes made from multilayer graphdiyne, a graphene-like crystal with a larger unit cell. Despite being nearly a hundred of nanometers thick, the membranes allow fast, Knudsen-type permeation of light gases such as helium and hydrogen whereas heavy noble gases like xenon exhibit strongly suppressed flows. Using isotope and cryogenic temperature measurements, the seemingly conflicting characteristics are explained by a high density of straight-through holes (direct porosity of ~0.1%), in which heavy atoms are adsorbed on the walls, partially blocking Knudsen flows. Our work offers important insights into intricate transport mechanisms playing a role at nanoscale.**

Porous membranes made on the basis of two-dimensional (2D) materials attract intense interest for their potential use in separation technologies (*1-11*). This interest is due to the fact that atomic-scale thickness implies very fast molecular permeation, as compared to conventional 3D membranes that exhibit flow rates scaling inversely proportional to the membrane thickness. To prove this ultimately fast permeability, 2D membranes with relatively large pores having the effective size $d_0$ larger than the kinetic diameter $d_k$ of sieved molecules have intensively been explored [for example, see (*1-3*)]. This regime is well described by the classical Knudsen theory and allows a moderate selectivity that arises from differences in thermal velocities of gases with different molecular weights $m$ (*1-2*). On the other hand, gas selectivity can greatly be improved using membranes with angstrom-scale pores of $d_0 \leq d_k$. In this case, molecules encounter substantial activation barriers for translocation through membranes, which leads to exponentially enhanced selectivity between gases having even marginally different $d_k$ (*4-5*). Unfortunately, the presence of activation barriers also implies an exponential suppression of flow rates (*4-5*). This tradeoff between permeability and selectivity is well known (*6*) and motivates the search for novel nanoporous materials with optimal tradeoff characteristics.

To create nanopores in 2D crystals, top-down fabrication is often utilized to introduce nanoscale defects in initially impermeable 2D materials (*1-5*). An alternative approach, perhaps more realistic in terms of applications, is bottom-up synthesis of thin nanoporous membranes such as, e.g., laminates made of 2D materials (*7-8*) and multilayer films of intrinsically porous crystals (*9-11*). In principle, such quasi-2D membranes with thicknesses comparable to the mean free path $\lambda$ of gas molecules should offer flow rates similar to those achievable using nanoporous strictly-2D crystals. However, mechanisms governing gas permeation and separation by quasi-2D membranes remain poorly understood (especially, experimentally) as, for example, they may differ from simple Knudsen and activated transport models. One of the candidate crystals for envisaged high-performance molecular-sieving membranes is graphdiyne, a carbon allotrope that has intrinsic triangular pores of a few angstroms in size (*12-13*). Its potential use in gas separation technologies has extensively been discussed through theory and simulations (*14-17*) but experimental assessment of its gas separation properties is still lacking.

In this work, we have investigated gas transport through graphdiyne-based nanoporous films, synthesized via coupling reactions of hexaethynylbenzene molecules (*18*). Details of the synthesis and characterization of the resulting material, using Raman and X-ray photoelectron spectroscopy, are provided in Supplementary Information and Fig. S1. As shown in Fig. 1 and Supplementary Fig. S2, the obtained graphdiyne films have a rather complex morphology and can conceptually be divided into two parts. One is a flat layer of ~90 nm in thickness, which consists of nanoscale multilayer graphdiyne crystallites aligned in plane. On top of this quasi-2D layer, crystallites grew mostly vertically and self-organized into a scaffold that can be viewed as interconnected nanometer-thick vertical walls or merged microwells of a submicron depth and a similar diameter (Fig. 1a). The scaffold provides sufficient mechanical support for the polycrystalline layer to allow free-standing membranes of several micrometers in diameter, which can withstand pressures up to $1 \times 10^5$ Pa (see below). Our extensive examination of the films using both scanning and transmission electron microscopy (TEM) revealed no defects larger than >50 nm in size (e.g., no cracks or tears). However, at the bottom of each microwell we observed small regions with vanishing thickness. These regions are separated by typical distances of ~100 nm as shown in Fig. 1b. Zooming into such thinnest regions, we estimated their size to be of the order of 10 nm. Unfortunately, the regions were found to be unstable under the electron beam exposure induced by high-magnification TEM imaging, as such we could not conclusively distinguish whether they were intrinsical straight-through holes or electron beam induced graphdiyne defects. Nonetheless, electron diffraction patterns taken near these regions confirmed ABC-stacked graphdiyne with its crystal structure shown in Fig. 1c, in agreement with the previous reports obtained from similar multilayer films (*19*).

To investigate gas transport through the graphdiyne-based films, we suspended them over micrometer-sized apertures etched in silicon-nitride/silicon wafers (insets of Fig. 2a). This was done following fabrication procedures described in *(20)*. The resulting membranes were placed in between two vacuum-tight chambers, one of which was filled with a gas under investigation while the other chamber was kept under high vacuum and connected to a mass spectrometer. For details of fabrication and measurement procedures, see Supplementary Information and Fig. S3.

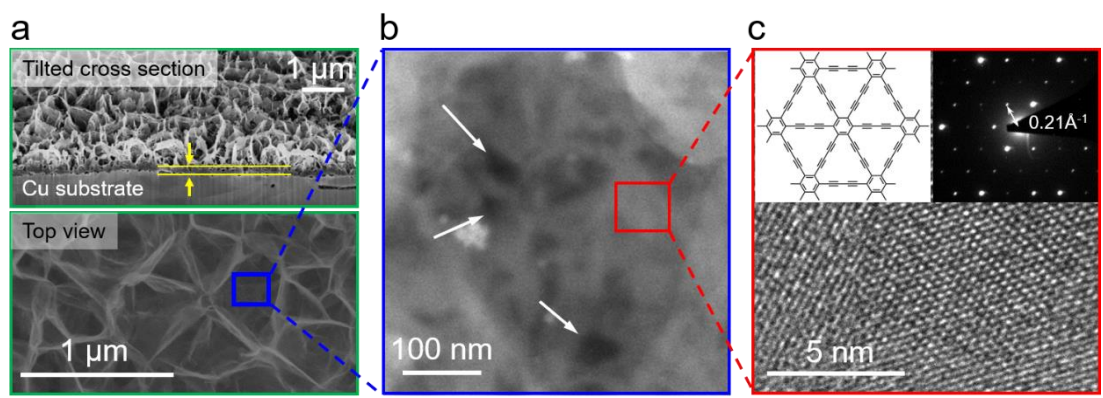

**Figure 1 | Graphdiyne-based membranes. (a)** Scanning electron microscope image of one of our membranes. Top panel shows cross-sections of the graphdiyne membrane, tilted by ~54° to show both the quasi-2D layer (also indicated by the yellow lines and arrows) and the vertical wall/ merged microwell structures on top. Bottom panel shows the top view of the membrane. **(b)** TEM image of the membrane. The thinnest regions at the bottom of microwells appear dark and are indicated by the arrows. **(c)** TEM image of a flat region near the bottom of a microwell (low panel). Top left: Schematic of monolayer graphdiyne's structure. Top right: Selected area electron diffraction pattern from the same region.

First, we studied permeation of various gases at room temperature $T$. Those included all nonradioactive noble gases (namely, $^3$He, $^4$He, Ne, Ar, Kr and Xe) and hydrogen isotopes $D_2$ and HD. Note $H_2$ was avoided because of a large fluctuating background usually arising in mass spectrometry for this particular isotope, which did not allow sufficient accuracy required for purposes of this report. The chosen gases provided a wide range of $m$ and $d_k$ values which allowed detailed characterization of molecular transport through our membranes. Examples of the measured gas flow rates $\Gamma$ as a function of the applied pressure $P$ are shown in Fig. 2a. As $\Gamma$ scaled linearly with $P$, gas transport through the graphdiyne-based films can be characterized by their permeability $\Gamma^* = \Gamma/P$. If the same apertures were covered by multilayer graphene (impermeable to gases (*21*)) or if a ~200 nm metal film was evaporated on top of graphdiyne post examination, no helium gas flow could be discerned within our detection accuracy of ~$10^{-14}$ mol s$^{-1}$. This corroborates that graphdiyne membranes were the only pathway for gas transport in our experiments. Furthermore, to calibrate our mass spectrometers with respect to different gases, we used "bare-hole" devices made in the same manner but without placing a graphdiyne film over the apertures. The reference devices exhibited $\Gamma^*$ approximately 1,000 times higher than that in the presence of graphdiyne-based membranes. This yields a porosity of ~0.1% for graphdiyne films. It is a remarkably high value, especially taking into account that the films easily withstood $P$ up to 1 × 10$^5$ Pa (higher $P$ were not tested). For comparison, nanometer-thick membranes made from graphene oxide (*7-8*), metal-organic frameworks, covalent-organic frameworks and zeolite (*9-11*) exhibited one to two orders of magnitude lower porosity.

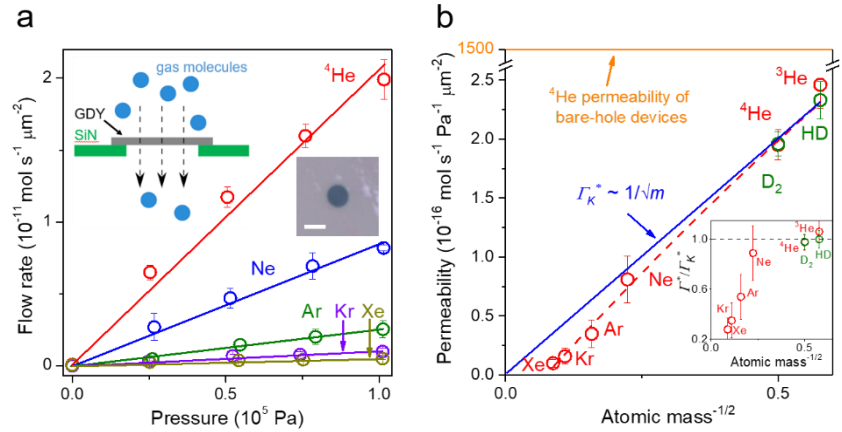

**Figure 2 | Gas permeation through graphdiyne-based membranes. (a)** Examples of the measured flow of noble gases through micrometer-sized membranes (symbols). Solid lines: Best linear fits to the data. Left inset: Schematic of our experimental setup. Right inset: Optical micrograph of one of our graphdiyne devices used in the experiments. The aperture is made in a 500 nm-thick silicon nitride (SiN) membrane and appears as a dark circle. It is covered by a suspended graphdiyne film (GDY). Scale bar, 2 μm. **(b)** Observed gas permeabilities at room temperature. Symbols are the experimental data with the error bars indicating Standard deviation (SD) using at least three different devices for each gas. The blue line shows the best fit by the Knudsen dependence using the data for light gases from $^3$He to Ne. Red curve: Guide to the eye. Inset shows the ratio of gas permeability to that from Knudsen dependence. For free molecular flows, the ratio is expected to be equal to one as indicated by the black dashed line.

Our results for $\varGamma^*$ are summarized in Fig. 2b. For light gases $^3$He, $^4$He, D$_2$, HD and Ne, the observed permeability can accurately be described by the Knudsen behavior $\varGamma^* \propto m^{-1/2}$ that is expected for free molecular flows ($\lambda \gg d_0 \gg d_k$). In contrast, heavy gases show pronounced deviations from the Knudsen dependence (inset in Fig. 2b). For example, Kr and Xe exhibited ~2.5 and 4 times lower rates, respectively, than those expected in the case of free molecular flows. This translates into the selectivity $S$ between $^4$He and Xe of ~20, well above $S \approx 5$ expected for the pure Knudsen flow. It is tempting to attribute the suppression observed for heavy atoms to sieving through the carbon mesh provided by the graphdiyne atomic structure (Fig. 1c), as widely discussed in the theoretical literature (*14-17*). Indeed, the crystal structure flaunts empty-space openings of nominally ~5.5 Å in diameter, if using the ball-and-stick model in Fig. 1c, and ~3.6 Å, according to the density functional calculations (*22*; Supplementary Fig. S4). Because this mesh size is comparable to $d_k$ of the studied gases, one can reasonably argue (*17, 22*) that the carbon mesh provides the necessary condition $d_k \approx d_0$ such that noble atoms with large kinetic diameters face a partial steric exclusion. However, this hypothesis contradicts the fact that HD and D$_2$ exhibited little difference in $\varGamma^*$ as compared to the same-mass but smaller $d_k$ $^3$He and $^4$He, respectively (hydrogen's $d_k \approx 2.9$ Å is relatively large, residing in between the kinetic diameters of Ne and Ar). Accordingly, if steric sieving was important, both hydrogen isotopes would show a notable suppression in $\varGamma^*$, in contrast to the experiment (Fig. 2b).

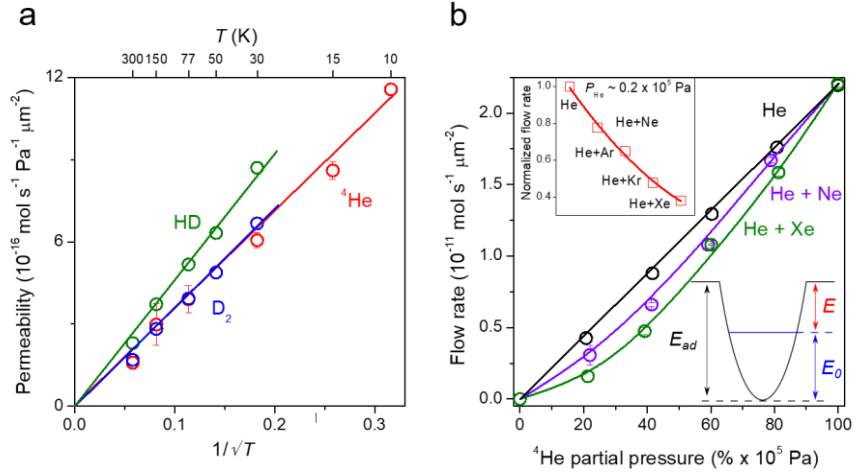

**Figure 3| Knudsen and non-Knudsen gas transport through nanoscale quasi-2D pores. (a)** Temperature dependence of gas permeability for light gases. Symbols: experimental data with the error bars indicating SD. Solid lines: best fits showing the Knudsen dependence. **(b)** $^4$He permeation as a function of its partial pressure within binary gas mixtures. The total pressure of the mixed gas is kept at $1 \times 10^5$ Pa. Solid curves: guides to the eye. Top inset: Helium flow rate at the partial pressure of $0.2 \times 10^5$ Pa with $0.8 \times 10^5$ Pa added by the other noble gases. The red solid line is a guide to the eye. Bottom schematic: Competition between the adsorption energy (black) and zero-point oscillations (blue) defines the binding energy (red) of adsorbed atoms.

To gain more information about the conflicting transport characteristics that cannot be explained by either the Knudsen flow or molecular sieving, we have studied temperature dependences of $\Gamma^*$ for helium and hydrogen gases (our setup does not allow low-$T$ measurements for gases with $m >$ 4). Within the accessible $T$ range from 300 K down to 10 K for helium and down to 30 K for hydrogen (the latter condenses at lower $T$), $\Gamma^*$ was found to vary as $\propto T^{-1/2}$ (Fig. 3a). This is the dependence expected for free molecular flow and consistent with the Knudsen behavior exhibited by all the light gases in Fig. 2b. The observation of the Knudsen $T$ dependence allows us to place an upper bound on the pore size $d_0$ in our graphdiyne films. Indeed, the Knudsen flow requires the condition $d_0 < \lambda \approx k_B T/(\sqrt{2}\pi d_k^2 P)$ to be satisfied over the whole range of $T$ and $P$ under investigation (*23*). Using our lowest $T$ of 10 K and highest applied $P$ of $1 \times 10^5$ Pa, we find that $d_0$ should be <5 nm. The validity of this analysis was crosschecked by measuring $T$ dependence for helium flowing through apertures of 30-50 nm in diameter. The relatively large apertures exhibited clear nonlinearities in permeability caused by transition from free molecular flow into the viscous regime, if either $P$ was increased or $T$ decreased. As for the lower bound on $d_0$, we first note that the pores cannot be smaller than the size of the graphdiyne carbon mesh discussed above. In addition, at cryogenic temperatures, permeating atoms and molecules can diffract at atomic- and nano- scale apertures because of a comparable de Broglie wavelength $\lambda_B = h/(3mk_B T)^{1/2}$ and aperture dimensions (*24*). For $^4$He at 10 K, $\lambda_B$ reaches ~4 Å. This value exceeds the kinetic diameters of the studied gases, which are given by the size of electron clouds around nuclei (*25*). If holes in our graphdiyne films were smaller than 1 nm, a contribution of quantum-mechanical diffraction should have been noticeable in the measured $T$ dependences or as an isotope effect for light gases (*26-27*). The purely Knudsen flow observed experimentally implies that $d_0$ should be considerably larger than $d_k + \lambda_B \approx 7$ Å. If we now take a theoretical perspective, our DFT calculations show that noble gases with $d_k$ close to $d_0$ should experience very large (~1eV) energy barriers (DFT simulations in Supplementary and Fig. S5; *ref. 5*). On the other hand, no sign of activated transport is noticeable in Fig. 3a implying that the barriers are less than a few meV, that is, comparable to or smaller than the thermal energy $k_B T$ at cryogenic $T$. This again suggests pores in our graphdiyne-based films to be larger than at least 1 nm. The above estimate 1 nm < $d_0$ < 5 nm is also consistent with the observed porosity and the described morphology of the graphdiyne films. Indeed, hole-like regions seen in our TEM images (i.e., nano-regions with vanishing thickness shown in Fig. 1b) are separated by a typical distance of ~100 nm. Therefore, the porosity of ~0.1% yields openings of ~3 nm in diameter within each region. All the above considerations allow us to conclude unambiguously that the graphdiyne films contain straight-through holes of a few nm in diameter, which are expected to provide Knudsen transport under all our experimental conditions, as indeed observed for light gases.

The inferred microstructure of the graphdiyne films seems to contradict the suppressed permeation observed for heavy noble gases. The clue to solve this final puzzle was found by measuring flow rates for binary mixtures of $^4$He with the other noble gases (Ne, Ar, Kr or Xe). The mixtures' $\Gamma$ exhibited profoundly nonlinear dependences as a function of the partial pressure of helium (Fig. 3b). For example, if $0.2 \times 10^5$ Pa of He was mixed with $0.8 \times 10^5$ Pa of Xe, the observed helium flow was more than twice slower than in the case of pure He at $0.2 \times 10^5$ Pa. This means that the presence of heavier noble gas atoms suppressed permeation of helium, in stark contrast to the well-known additive behavior for gas mixtures (especially, for inert gases). In other words, if mixed, noble gases no longer flow independently through graphdiyne pores even at room $T$. To the best of our knowledge, such interaction between flows of noble gases has never been observed before. We attribute this phenomenon to adsorption of heavy atoms on inner walls of straight-through holes in the graphdiyne films. As the holes are only ~10 times larger in size than typical $d_k$ and are expected to have a finite length (at least of several interlayer distances or a few nm), incident noble gas atoms would see the nanopores as partially filled with a heavy gas. This effectively reduces $d_0$ and leads to a suppressed flow of helium. For the case of pure heavy gases like Kr and Xe, this partial filling should also play a role and can be interpreted as either partial blockade of pores by same-gas atoms or their longer translocation times. This explanation agrees well with the progressively weaker interaction effect observed for lighter noble gases that are expected to exhibit weaker adsorption (top left inset of Fig. 3b). The weaking of the interaction effect with decreasing $m$ also suggests that $m$ is the defining parameter. To this end, we note that a comprehensive description of gas adsorption phenomena normally necessitates the inclusion of zero-point oscillations (*22, 28-29*). As illustrated in the bottom right inset of Fig. 3b, zero-point energies $E_0$ of adsorbed light molecules compete with adsorption energies $E_{ad}$ and effectively reduce the binding energy $E = E_{ad} - E_0$. Because $E_0$ varies as $m^{-1/2}$, $E$ is expected to decrease for light atoms and can even become negative (that is, no sticking of light gases to graphdiyne occurs). Judging from Fig. 3a, $E$ for helium and hydrogen should be less than a few meV.

To conclude, in comparison with the previously reported atomically-thin membranes with nanopores obtained by top-down fabrication, our quasi-2D membranes exhibit similar selectivities combined with high flow rates thanks to an extremely high pore density (~$10^{10}$ cm$^{-2}$).

Supplementary Fig. S6 also suggests our membranes provide a better permeability-selectivity performance beyond the existing trade-off bounds. Unexpectedly, adsorption plays a completely different role in these quasi-2D membranes as compared to 2D membranes such as, e.g., in perforated graphene. Molecules adsorbed on graphene can easily move in-plane (*30-31*), which enhances permeation by many orders of magnitude. In contrast, adsorption of heavy atoms on internal surfaces of the graphdiyne-based membranes reduces permeation. Moreover, the inner-pore adsorption gives rise to a counterintuitive effect of interacting flows of supposedly non-interacting, inert gases. No noticeable gas transport through the intrinsic mesh within graphdiyne's crystal structure has been evidenced, due to their small effective diameter of < 4 Å which yields high energy barriers (Supplementary for DFT simulations) and/or the non-aligned intrinsic meshes from adjacent atomic layers in ABC stacked multi-layer graphdiyne that blocks gas flows. To this end, carbon allotropes with larger unit cells could be better candidates for gas separation, if their mechanical stability can be achieved.

**Supplementary Information**

**Membrane synthesis and characterization**

Graphdiyne-based films were synthesized via cross-coupling reactions of the monomer 1,2,3,4,5,6-hexaethynylbenzene (HEB) (*1*). In brief, first, HEB was synthesized by adding tetrabutylammonium fluoride to tetrahydrofuran solutions of hexakis[(trimethylsilyl)ethynyl]benzene at 0 °C and used after reaction time of 10 minutes. Next, HEB was dissolved in pyridine and was added slowly (in 8 hours) to a mixture of treated copper foils immersed in pyridine at 110 °C under argon atmosphere, with the ratio of HEB quantity to area of copper foils ~1 µg/cm$^2$. The mixture was kept at cross-coupling reaction conditions (argon atmosphere; 110 °C) for 64 hours and after that, graphdiyne-based films were grown on the surface of copper foils. Note that the entire process should avoid solution/reactant in contact with oxygen (in air) and light. Consequently, the graphdiyne-based membranes were rinsed using heated acetone and N, N-dimethylformamide to remove any residues (e.g. unreacted monomers and oligomer; solvents), and were dried under argon.

The as-synthesized graphdiyne-based membranes were characterized by X-ray photoelectron spectroscopy (XPS) and Raman spectroscopy (Fig. S1). The results are in consistent with those reported and indicate the presence of graphdiyne structures (*1*). In Fig. S1a, the area ratio of peaks sp/sp$^2$ is close to 2, in agreement with the graphdiyne structure of benzene rings connected by double -C≡C- links. The presence of C-O and C=O are attributed to air adsorptions. In Fig. S1b, the peak at 1373 cm$^{-1}$ and 1556 cm$^{-1}$ are attributed to the breathing vibration and stretching modes of sp$^2$ carbons in aromatic rings, respectively; the peaks at 1925 cm$^{-1}$ and 2170 cm$^{-1}$ correspond to the vibration of conjugated -C≡C-C≡C- bonds.

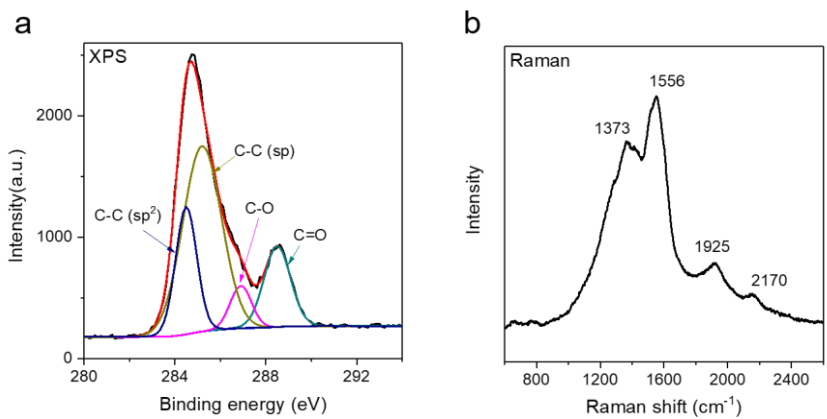

**Figure S1 | Graphdiyne membrane characterization,** using **(a)** XPS and **(b)** Raman spectroscopy techniques. The spectrum peak position assignment in (a) and (b) are in accordance with ref., respectively.

**Morphology characterization**

Scanning electron microscope (SEM) images were obtained using Zeiss Ultra Scanning Electron Microscope. By tilting the SEM stage at an angle, the interconnected nanometer-thick vertical walls and the microwell structures can be clearly seen (Fig. S2a, S2b). Cross-sectional images of graphdiyne membranes were taken on Zeiss Cross-beam SEM/FIB system. To expose the cross sections, an area of ~10 × 2 µm$^2$ trench was removed using focused Ga ion-beam (Fig. S2c). Notably, morphology of graphdiyne films near the trench appears bright compare to its original appearance, as a result of re-deposition during ion milling process (Fig. S2d). Transmission electron microscope (TEM) images were obtained using either FEI Titan G2 80-200 or Thermo Scientific Talos F200S at 200 keV.

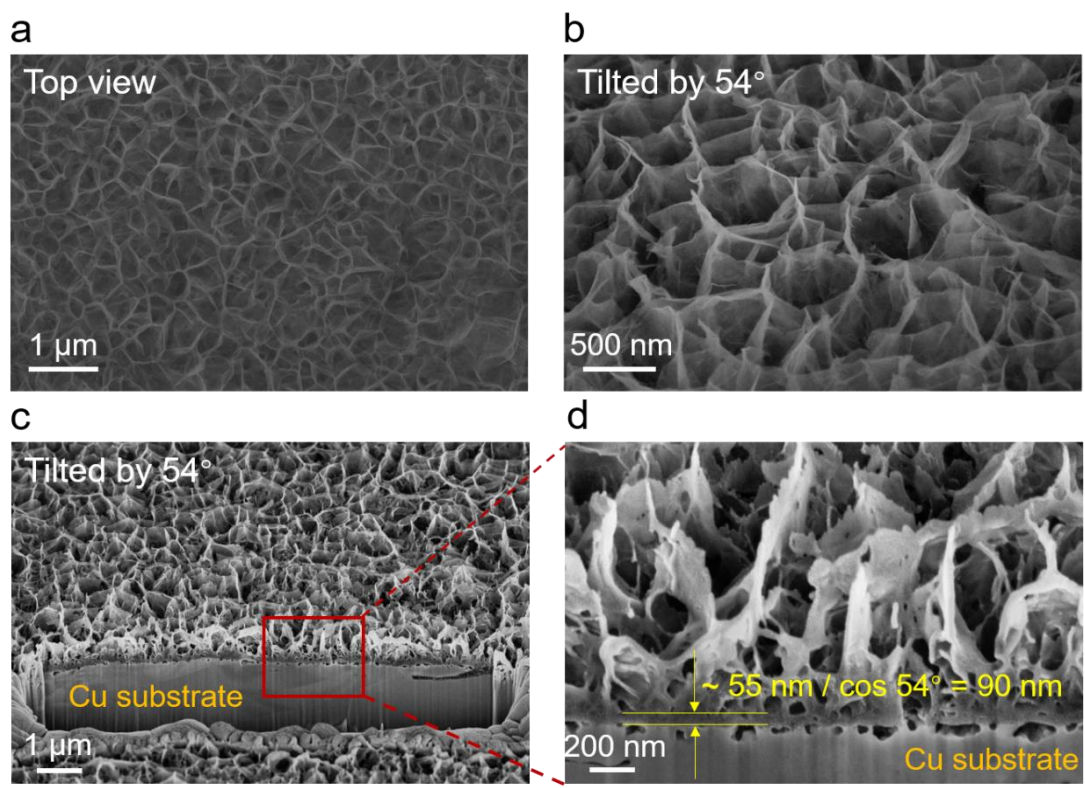

**Figure S2 | Morphology characterization using electron microscopy. (a)** and **(b)** SEM images of as-prepared graphdiyne membranes. **(c)** and **(d)** SEM images at cross-sections prepared by focused ion beam.

**Gas transport device fabrication and measurements**

Fabrication procedures of our gas transport devices are as follows. First, a free-standing silicon nitride membrane with a micrometer-sized aperture in the center was prepared using standard photolithography and reactive ion etching techniques, following the procedures described in *ref. 2*. Next, graphdiyne-based membranes were transferred on top of the silicon nitride membrane using wet transfer methods (*3*) to cover the aperture there. To measure gas permeation through graphdiyne membranes, we use experimental set-up as shown in Fig. S3a. In short, the devices were sandwiched between two He leak tight (leak rate < $10^{-14}$ mol s$^{-1}$) chambers. One chamber is filled with gases under investigation (or their binary mixtures for the measurements shown in Fig. 3b) at variable pressures up to 1 bar, while the other chamber is kept at vacuum and is connected to either a mass spectrometer; or a helium leak detector which is sensitive to detect flow rates for gases with molecular mass *m* = 3, 4.

To calibrate gas permeability measured by the helium leak detector, we note that it allows internal calibration for the flow rate of $^4$He. As a control experiment to demonstrate the accuracy of calibration, we tested the permeation of $^4$He through a 1-μm (in diameter) aperture in silicon nitride. The leak detector correctly measured permeability $\Gamma^*_{4He}$ that approaches (error < 10%) its theoretical value of $\frac{8}{3N_A} \frac{d_0}{2L} \frac{1}{\sqrt{2\pi mkT}}$, where $L$ is the length of the aperture. In these measurements, the pressure applied is kept < 0.1 bar to ensure the Knudsen condition (mean free path $\lambda$ > aperture dimension $d_0$, as discussed in the main texts) is fulfilled. For the calibration of other gas species, a correction factor was calculated from the ratio between their measured permeability and that of $^4$He (more specifically, $\Gamma^*_{4He} \times \sqrt{\frac{m_{4He}}{m_0}}$, with $m_{4He}$ and $m_0$ the atomic mass of $^4$He and gases under investigation, respectively) through the aperture. Similar procedures were applied to the calibration of gas permeation measured by the mass spectrometer, using the same 1-µm aperture.

For measuring gas permeation at cryogenic temperatures, we use a home-made constant flow cooling system as shown in Fig. S3b. The cooling chamber was connected to a liquid helium reservoir via a transfer tube. The transfer tube has a concentric geometry and was equipped with a needle valve at the reservoir end of the tube. By pumping out the cooling chamber, the liquid cryogen expands through the needle valve into cold gas that flows in the center layer of the tube towards the sample chamber region. Surrounding the center layer, an outer layer of less cold gas (compare to that in the center layer) flows in reverse direction from the sample region to the exhaust (to a helium recovery line) and acts as a radiation shield. The flow rate of cryogen (and hence cooling rate) was controlled by adjusting the pumping speed of the vacuum using another valve between the pump and the transfer tube. To make holding at a constant temperature easier, a heater was used to provide heating. The temperature was measured using a calibrated thermocouple. When running this cooling setup, the helium leak detector background rose to up to $10^{-13}$ mol s$^{-1}$. This 1-2 orders of magnitude higher background than that measured under ambient conditions was a consequence of He leakage through the fittings of the pipe-work into the atmosphere which was picked up by the sensitive leak detector. This background was however still at least one order of magnitude smaller than our smallest signals and thus has negligible influence to the accuracy of measurements.

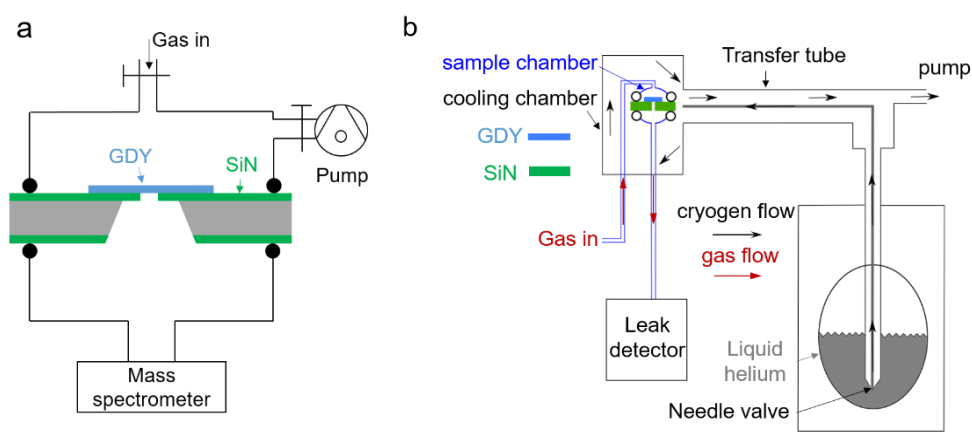

**Figure S3 | Schematics of gas permeation measurements set-up: (a)** For measurements at non-cryogenic conditions. Black circles represent rubber O-rings for sealing. **(b)** For cryogenic temperature measurements. Indium seals (black empty circles next to the SiN wafer) were used instead of rubber O-rings which might fail at low temperatures.

**DFT simulations**

For the density functional theory-based first-principles calculations, the projector augmented wave (*4*) method was used in order to portray the pseudopotentials of C atom and the noble gas elements as implemented in Vienna Ab-initio Simulation Package (VASP) (*5*). The exchange-correlation potential was taken into account by considering the local density approximation (LDA) within the Perdew-Burke-Ernzerhof (PBE) form (*6*). For the geometry optimizations a kinetic energy cutoff of 500 eV was used for the plane-wave basis. The convergence criterion of the total force on each atom was reduced to $10^{-5}$ eV/Å and the convergence criterion of the energy was set to $10^{-6}$ eV.

The primitive unit cell of graphdiyne structure is composed of 18 C atoms and it was optimized using a *k*-point mesh of 18 × 18 × 1. In order to simulate the propagation of noble gas atoms through the hole of the graphdiyne structure, a super cell containing 72 C atoms was used. We checked that this was sufficiently large to avoid inter simulation cell interaction. The initial distance of a noble gas atom to the basal plane of graphdiyne was taken to be 8 Å and the vertical position of the noble gas atom was changed by 0.5 Å steps and then the total energy was calculated at each step. Note that while the rigid structure refers to the case when all the carbon atoms are frozen, the relaxed case refers to fully optimization of the whole structure as gas atoms diffuse through the lattice rings in graphdiyne.

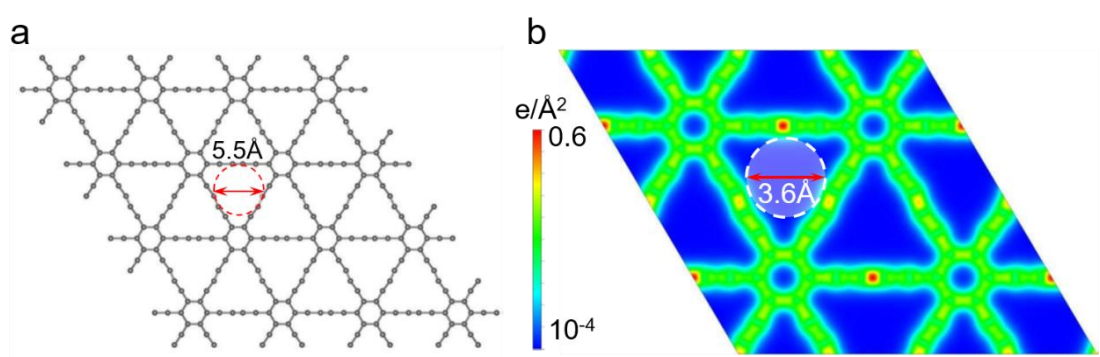

**Figure S4 | Lattice rings in graphdiyne. (a)** Schematic of a monolayer graphdiyne using ball-and-stick model. The grey balls represent carbon atoms. **(b)** Graphdiyne electron density predicted by density functional theory. The electron density (in electrons per Å$^2$) is integrated along the direction perpendicular to graphdiyne.

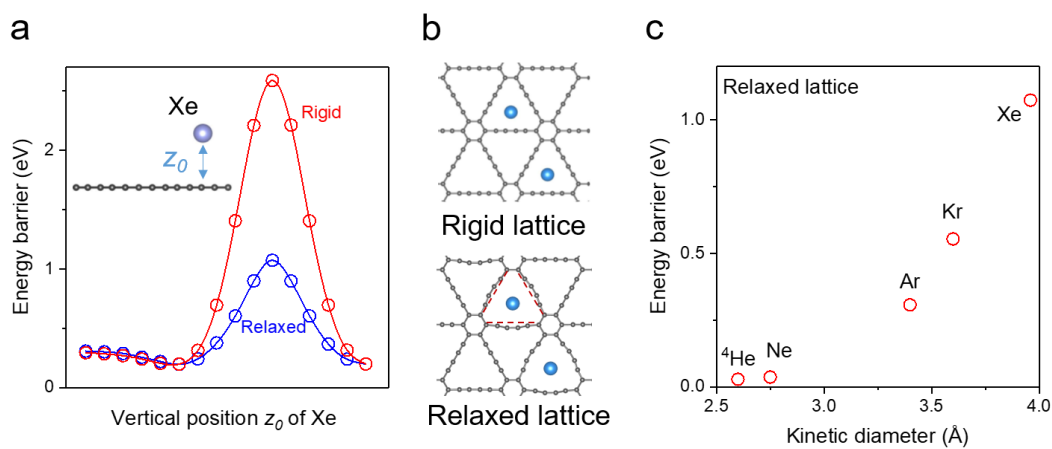

**Figure S5 | Activated gas transport through graphdiyne lattices. (a)** Energy profiles of Xe atoms through rigid and relaxed graphdiyne lattice rings, as a function of the vertical position $z_0$ (defined as indicated by the inset figure) between Xe and graphdiyne plane. **(b)** Schematic of rigid and relaxed lattice when Xe atoms locate in the lattice plane, i.e. $z_0 = 0$ in (a). **(c)** Transport barriers of various noble gases through relaxed graphdiyne lattice.

**Comparison of permeability-selectivity tradeoff with other gas selective membranes**

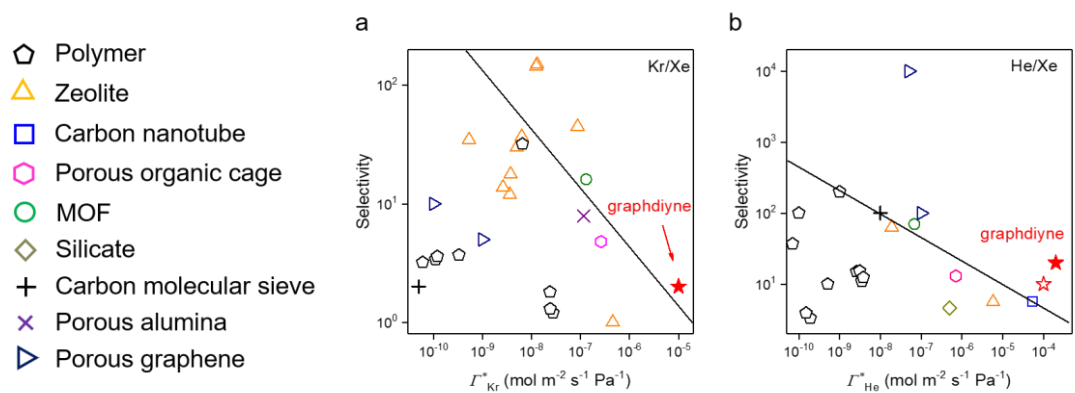

**Fig. S6 | Selectivity – permeability upper bound relations for various membranes. (a)** Kr-Xe separation. **(b)** He-Xe separation. Solid black lines indicate the state-of-art boundary. Filled red star represents the separation performance from the data of single component gas transport through graphdiyne-based membranes; for binary mixtures, both the permeability and selectivity are corrected by about 50%, as shown by the empty red star in (b). The literature data include membranes made from polymers (*7-10*), zeolites (*11-17*), carbon nanotubes (*18*), porous organic cages (*19*), metal organic frameworks (MOF) (*20-21*), silicate (*22*), carbon molecular sieves (*23*) and porous alumina (*24*). The figure also includes results from previously reported porous graphene. For those of perforated graphene with single-nanopores (*25*), we assume their pore density of $10^{12}$ cm$^{-2}$ (which is the highest density currently achievable by top-down methods) without loss of single-pore functionality.